\documentclass[twocolumn]{aastex62}
\usepackage{outlines}

\graphicspath{{./}{figures/}}

\received{April 1, 2019}
\submitjournal{Acta Prima Aprilia}

\shorttitle{The Climate of Westeros}
\shortauthors{Paradise et al.}

\begin{document}

\title{The Long Night: Modeling the Climate of Westeros}

\correspondingauthor{Adiv Paradise}
\email{paradise@astro.utoronto.ca}

\author[0000-0001-6774-7430]{Adiv Paradise}
\affil{Department of Astronomy \& Astrophysics ,University of Toronto \\
50 St. George Street, Toronto, ON, M5S 3H4, Canada}
\affil{Centre for Planetary Sciences, University of Toronto, Scarborough \\
1265 Military Trail, Toronto, ON, M1C 1A4, Canada}

\author{Alysa Obertas}
\affil{Department of Astronomy \& Astrophysics ,University of Toronto \\
50 St. George Street, Toronto, ON, M5S 3H4, Canada}
\affil{Centre for Planetary Sciences, University of Toronto, Scarborough \\
1265 Military Trail, Toronto, ON, M1C 1A4, Canada}
\affil{Canadian Institute for Theoretical Astrophysics, University of Toronto \\
60 St. George Street, Toronto, ON, M5S 3H8, Canada}

\author[0000-0002-7296-6547]{Anna O'Grady}
\affil{Department of Astronomy \& Astrophysics ,University of Toronto \\
50 St. George Street, Toronto, ON, M5S 3H4, Canada}
\affil{Dunlap Institute for Astronomy \& Astrophysics, University of Toronto \\
50 St. George Street, Toronto, ON, M5S 3H4, Canada}

\author[0000-0003-1090-1673]{Matthew Young}
\affil{Department of Astronomy \& Astrophysics ,University of Toronto \\
50 St. George Street, Toronto, ON, M5S 3H4, Canada}
\affil{Dunlap Institute for Astronomy \& Astrophysics, University of Toronto \\
50 St. George Street, Toronto, ON, M5S 3H4, Canada}
\begin{abstract}

Many previous authors have attempted to find explanations for Westeros's climate, characterized by a generally moderate, Earth-like climate punctuated by extremely long and cold winters, separated by thousands of years. One explanation that has been proposed is that the planet orbits in a Sitnikov configuration, where two equal-mass stars (or a star and a black hole) orbit each other on slightly eccentric orbits, and the planet moves along a line through the barycenter perpendicular to the primaries' orbital plane \citep{Freistetter2018}. We modify an intermediate-complexity GCM to include the effects of such an orbit and integrate it for thousands of years to determine whether such an orbit can a) be habitable and b) explain the climatic variations observed by the inhabitants of Westeros, in both double-star and star-black hole configurations. While configurations with low primary eccentricity and initial conditions that permit only small excursions from the ecliptic plane are habitable, these orbits are too stable to explain Westerosi climate. We find that while orbits with more bounded chaos are able to produce rare anomalously long and cold winters similar to Westeros's Long Night, huge variations in incident stellar flux on normal orbital timescales should render these planets uninhabitable. We note that the presence of an orbital megastructure, either around the planet or the barycenter, could block some of the sunlight during crossings of the primaries' orbital plane and preserve Westeros's habitability. While we find that bounded chaotic Sitnikov orbits are a viable explanation for Westeros's Long Night, we propose that chaotic variations of the planet's axial tilt or semimajor axis, potentially due to torques from nearby planets or stars, may be a more realistic explanation than Sitnikov orbits.

\end{abstract}


\section{Introduction} \label{sec:intro}

A major concern among Westerosi people, especially those living at high latitudes, is the belief that Winter is Coming, at some unknown time in the future \citep{Martinbook}. Here we distinguish Westerosi Winters (with a capital `W') from winters (with a lowercase `w'), such that `winter' refers to normal decreases in temperatures experienced in Westeros frequently, every few months to years, and `Winter' refers to the Long Night, a catastrophic life-threatening event characterized by cold lasting a generation, kings freezing on their thrones, and White Walkers advancing and threatening the Kingdoms of Men \citep{Martinbook}.

Westerosi Winters are believed unpredictable by the maesters and pose a severe threat to health and safety in Westeros. This makes understanding their cause a major priority. Discovering their geophysical cause may even lead to the ability to predict when the next Winter will occur, allowing the Seven Kingdoms ample time to set aside their squabbles and prepare food stores and armies to face the coming Winter. Understanding the climate of Westeros also provides a case study for a habitable climate whose time-evolution and orbital environment is very different from Earth's.

A great deal of work has already gone into explaining the climate of Westeros. \citet{Kostov2013} proposed that Westeros could inhabit a P-type circumbinary system, where the planet orbits both stars in a binary, noting that the orbit could vary chaotically, leading to variable and unpredictable lengths of Westerosi winters. \citet{Freistetter2018} however noted that there is no evidence of multiple suns in Westerosi skies, and that the variations in a P-type circumbinary orbit are too small and too regular to explain the severity, duration, and infrequency of the Long Night, instances of which can be separated by as many as 8,000 years \citep{Martinbook}. They proposed that Sitnikov orbits, which can vary much more chaotically, are instead a more plausible explanation, noting that the single-sun problem can be resolved by having one of the primaries be a black hole. \citet{Freistetter2018} however did not actually calculate how the amount of incident sunlight, or insolation, would vary throughout a Sitnikov orbit, or make any attempt to actually model the climate's response to such an orbit.

Sitnikov orbits, a special case of the restricted three-body problem, are characterized by an eccentric stellar binary with the planet orbiting on a straight line through the system's barycenter, such that the axis of motion is perpendicular to the primaries' semimajor axis \citep{Sitnikov1961}. The duration of excursions away from the plane is chaotic, such that there exist initial conditions which lead to any integer sequence of primary orbits between ecliptic crossings, though finding those initial conditions may be difficult \citep{Moser1973}. Sitnikov orbits can be stable (with small chaotic variations), bounded but chaotic (with larger variations), and unbounded and chaotic (where the planet is eventually ejected) \citep{Kovacs2007}. Some orbits in particular can be ``sticky", such that a planet appears stable for a large number of orbits before becoming unbound \citep{Dvorak1998}. We note that the impact of three-body chaos on planetary climate is habitability is also explored in considerable depth in \citet{cixinliu2014}.

While our work focuses on equal-mass binaries, approximates the planet as a massless particle, and restricts the planet to orbiting on the perpendicular axis, other studies have shown that these properties of Sitnikov orbits also apply to systems with unequal-mass binaries \citep{Perdios1987}, planets with mass and influence on the parent binary \citep{Dvorak1997}, and systems where the planet's orbital motion is not exactly on the perpendicular axis \citep{Perdios2012}.

Examining astrophysical explanations for Westeros's unusual climate is interesting from an exoplanet science point of view. To date, a huge number of exoplanets has been discovered \citep{Mullally2017}. Within this sample we find an enormous diversity of planets, many of which defy easy explanations. The very first exoplanet ever discovered, PSR B1620-26b, belongs to a binary system with a pulsar and white dwarf \citep{Backer1993}. The second exoplanet ever discovered, 51 Pegasi b, is a hot Jupiter, orbiting much closer to its star than Mercury orbits our own Sun \citep{Mayor1995}. While it is not clear how a Sitnikov system could form, the large number of planets discovered and expected to be discovered makes it worthwhile to explore the fringes of the exoplanet parameter space.

Another motivation for studying this problem is the similarity between the Long Night and snowball climates, where the oceans are completely covered by sea ice \citep{pierrehumbert05,Abbot2014,Hoffman2017}. These climates are stable at the same greenhouse gas concentrations and insolations as temperate climates, and the two states are separated by sharp transitions to a runaway ice-albedo feedback \citep{Budyko1969,Sellers1969,icealbedo}. This has happened at least twice on Earth \citep{Hoffman2002,dropstones,Tajika2007}, and is expected to happen on at least some Earth-like exoplanets \citep{Menou2015,Abbot2016,Paradise2017}. It is not expected to be very common however on tidally-locked planets \citep{Checlair2017}. Westeros therefore represents an opportunity to investigate another regime in which snowball glaciations may be possible, and orbital chaos may represent another mechanism for triggering them. 

\section{Methods} \label{sec:methods}

\subsection{Sitnikov Orbit}\label{sec:orbit}

We assume the orbits of the primaries are Keplerian and unaffected by the planet. The equation of motion for a massless particle on a Sitnikov orbit is given by \citet{Sitnikov1961}:
\begin{equation}
    \ddot{z}(t) = -\frac{2GMz(t)}{\left(r^2(t)+z^2(t)\right)^{3/2}}
\end{equation}
where $G$ is the gravitational constant and $M$ is the combined mass of the primaries. The distance of the primary from the barycenter, $r(t)$,  can be easily calculated through
\begin{equation}
    r(t) = a\frac{1-e^2}{1+e\cos\nu(t)}
\end{equation}
where $e$ is the primary's eccentricity, $a$ is the primary's semimajor axis, and $\nu$ is the primary's true anomaly. We calculate $\nu$ from the primary's mean anomaly and eccentric anomaly, the latter of which is calculated with a Newton-Raphson iterator, run until the result is converged to within machine error ($<10^{-15}$). We integrate the orbit of the planet using a 4th-order Yoshida integrator \citep{Yoshida1990}. We integrated several configurations for several thousand orbits with both a Yoshida integrator and a leapfrog integrator, and found negligible differences in most cases, and in cases where the planet was ejected, the main difference was the direction in which it was ejected. We therefore conclude that with more than $10^4$ timesteps per orbit, a 4th-order Yoshida integrator has sufficient accuracy for our purposes. 

\subsection{Planetary Rotation}\label{sec:rotation}

\begin{figure}
\plottwo{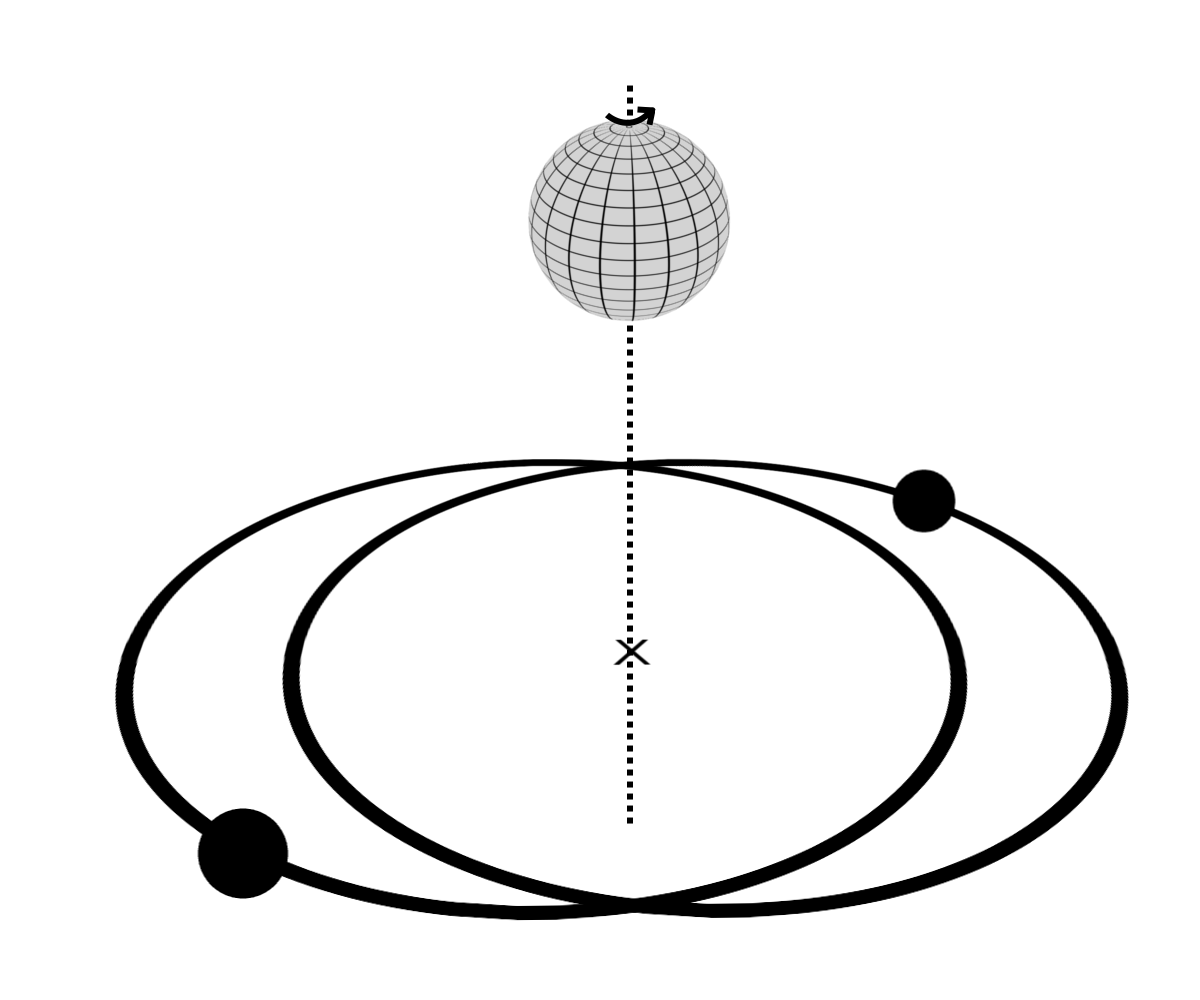}{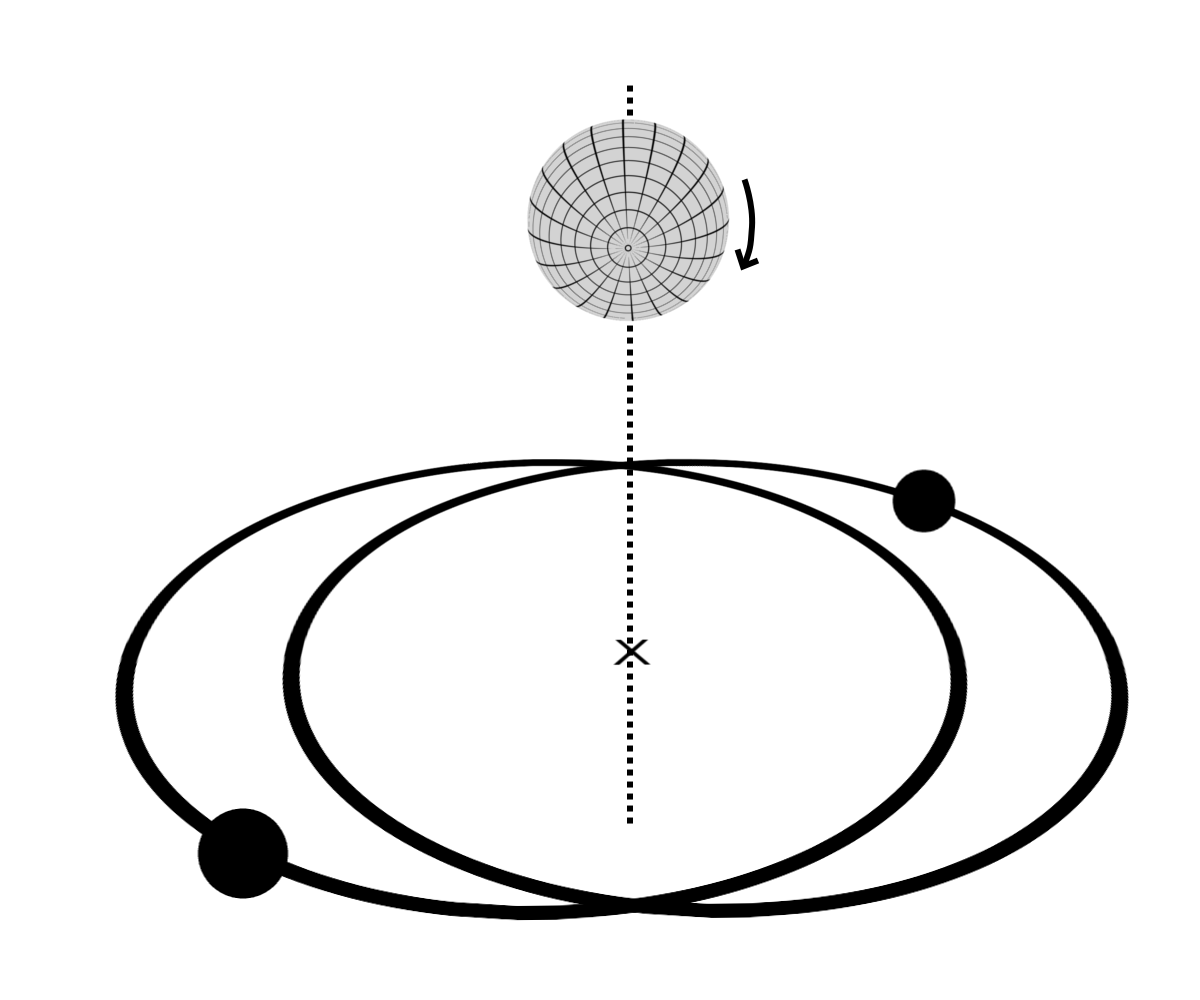}
\caption{Diagrams of the Sitnikov orbital configuration with the two rotation states we consider. The spinning-top configuration (left) has the planet's rotation axis parallel to the axis of motion, while the tennis-ball configuration (right) has the planet's rotation axis perpendicular to both its own semimajor axis and the primaries' semimajor axis.\label{fig:rotation}}
\end{figure}

 We assume two possible geometries for the planet's rotation relative to the direction of motion, which we term the spinning-top and tennis-ball configurations, shown in \autoref{fig:rotation}. In the spinning-top configuration, the planet's rotation axis is aligned with the direction of motion (similar to a spinning top gently levitating above a magnetic base), such that day and night progress from East to West, and only the declination of the primaries changes as the planet moves throughout its orbit. This is analogous to having 90$^\circ$ obliquity in a Keplerian orbit. In the tennis-ball configuration, the planet's rotation axis is perpendicular to the direction of motion (the semimajor axis). This is analogous to the low obliquity of most Solar System planets, and is similar to the rotation of a tennis ball tossed in the air.

We choose to orient the rotation axis of the tennis-ball configuration perpendicular to both the planet's semimajor axis and the primaries' semimajor axis, such that in the case of primaries on eccentric orbits, day and night progress longitudinally when the planet is above or below the plane, and the primaries are at low declinations at apoapse, where they spend most of their time. The formulae for the stellar angles in each configuration are given in \autoref{appendix:angles}

\subsection{Climate Model Configuration}

To model the climate of Westeros, we use PlaSim, an intermediate-complexity 3D global climate model described in \citet{Fraedrich2005}. PlaSim solves a set of primitive equations on a discretized grid using a spectral transform method. PlaSim represents the ocean as a 50-meter mixed slab without horizontal advection (as in, heat is transported only through diffusion), models sea ice thermodynamically (ice forms when it is cold enough to do so), and couples the atmosphere to the land and ocean through latent and sensible heat transport (shallow and deep convection and precipitation/evaporation), drag, and radiation. PlaSim incorporates a 3-band radiation model, with two shortwave bands and one longwave band. Absorption is by water, CO$_2$, and ozone, while scattering is computed based on the total mass of the atmosphere and cloud optical depth. Clouds are computed diagnostically based on critical relative humidities for each vertical layer, and precipitation includes both convective and large-scale precipitation, and models all three phases of water---precipitation can fall as rain or snow, and water can be re-evaporated out of precipitation before it reaches the ground \citep{Fraedrich2005}. 

PlaSim has been used to study a number of Earth-like climates, including snowball climates, where sea ice extends to the equator \citep[e.g.][]{lucarini2010,Boschi2013,Paradise2017}, high-obliquity climates \citep{Linsenmeier2015}, tidally-locked planets \citep{Checlair2017,Abbot2018}, and planets undergoing a moist greenhouse \citep{Gomez-Leal2019}. To capture the unique climate dynamics of a planet on a Sitnikov orbit, we introduce modifications to PlaSim to allow for chaotically-evolving incident stellar flux and multiple suns in the sky. We integrate the positions of the primaries and the planet each timestep, and then compute the solar zenith angle independently for each primary. Because some of the scattering and absorption parameterizations depend on the angle of sunlight propagating through the atmosphere, we compute shortwave radiative transfer independently for each primary, and then combine the fluxes from each at the layer boundaries \citep{GPlaSim}.

We run PlaSim in its T21 resolution, corresponding to 32 latitudes and 64 longitudes ($5.6^\circ\times5.6^\circ$ at the equator), with 10 vertical levels spaced mostly-linearly in pressure. We run our Yoshida integrator for 10$^4$ years for several hundred different initial conditions, with varying initial mean anomalies, eccentricities, and initial $z$, and then from the output select 2 initial conditions that yield bounded chaotic orbits, with infrequent extremely cold and long winters. We test whether a spinning-top configuration (see \autoref{sec:rotation}) can be habitable at the initial elevations necessary for significant chaotic variations by using generic initial conditions ($e=0.1$, initial mean anomaly $M=0.8$) at several initial elevations, ranging from 0.3 to 0.8 AU. Because the spinning-top configuration at initial elevations much beyond this range plunge the antistellar hemisphere into darkness for a significant portion of the orbit, while delivering weakened light to the near hemisphere, the spinning-top configuration plus a strongly-chaotic orbit results in model-crashing cold temperatures during each orbit. We cannot rule out the role of White Walkers on such planets, but to be safe we use the tennis-ball configuration for the simulations with chaotic orbits. Because the chaotic configurations lead to increases in stellar flux at ecliptic-crossing of a factor of 5--6 relative to maximum excursion above the plane, in some models we place an upper limit on the flux that the planet can receive, choosing 1400 W/m$^2$. We run each simulation for 2000 years, with a timestep of 45 minutes. 

This approach of first only integrating the orbits to find promising initial conditions and then re-integrating during the climate simulation itself turned out to be a mistake, since the chaotic nature of Sitnikov orbits and the finite precision of machine arithmetic meant that the final orbits differed often significantly from the initial calculations, with 2 of our 4 2000-year simulations being overcome by White Walkers and not emerging from the Long Night within the timescale of the simulation (may Jon Snow rest in peace). Due to time constraints, we were unfortunately not able to run more simulations to rectify this problem, but note that a more intelligent approach would have been to pre-calculate the orbits to find promising initial conditions, and then feed the calculated orbit into PlaSim as a prescribed time series, interpolating between the larger timesteps of the initial calculation rather than re-integrating everything. A hybrid method could also be possible, where the prescribed timesteps, of which there may be one per day or fewer, are used to `guide' the integration of the non-prescribed timesteps. 

\begin{figure}
\plotone{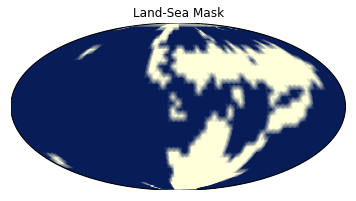}
\caption{Land-Sea mask for our Westeros-like model setup, shown here in a Mollweide projection (such that the entire planet is shown). Oceans are shown as blue/dark, while land is beige/light. Note that Sothoryos, Ultos, and the Sunset Sea are completely speculative, as are latitudes of various locations.\label{fig:geography}}
\end{figure}

Our Westeros planet is mostly Earth-like, with a surface atmospheric pressure of 1 atm, 360 $\mu$bars of CO$_2$, and a rotation rate of 24 hours. Due to the absence of modern meteorology equipment among the Maesters of Westeros, when it comes to the true surface pressure and pCO$_2$, we know about as much as Jon Snow, which is to say, nothing \citep{Martinbook}. For simplicity, here we assume them to be Earth-like. In the name of making the model as realistic as possible, however, we attempt to replicate the known land distribution of Westeros, shown in \autoref{fig:geography}. This land distribution is based on published maps of the known world \citep[e.g][]{officialmaps,atlasmap,gmapwesteros}, but we note that because the world is not fully explored \citep{martininterview}, the southern hemisphere and Sunset Sea are purely speculative. We note that we assume that the North beyond the Wall extends to the north pole, such that the Arctic pole is land, rather than ocean. We do not however expect large differences between our map and reality to have a large effect on the model's outcomes.

\section{Results}

\begin{figure*}
\plotone{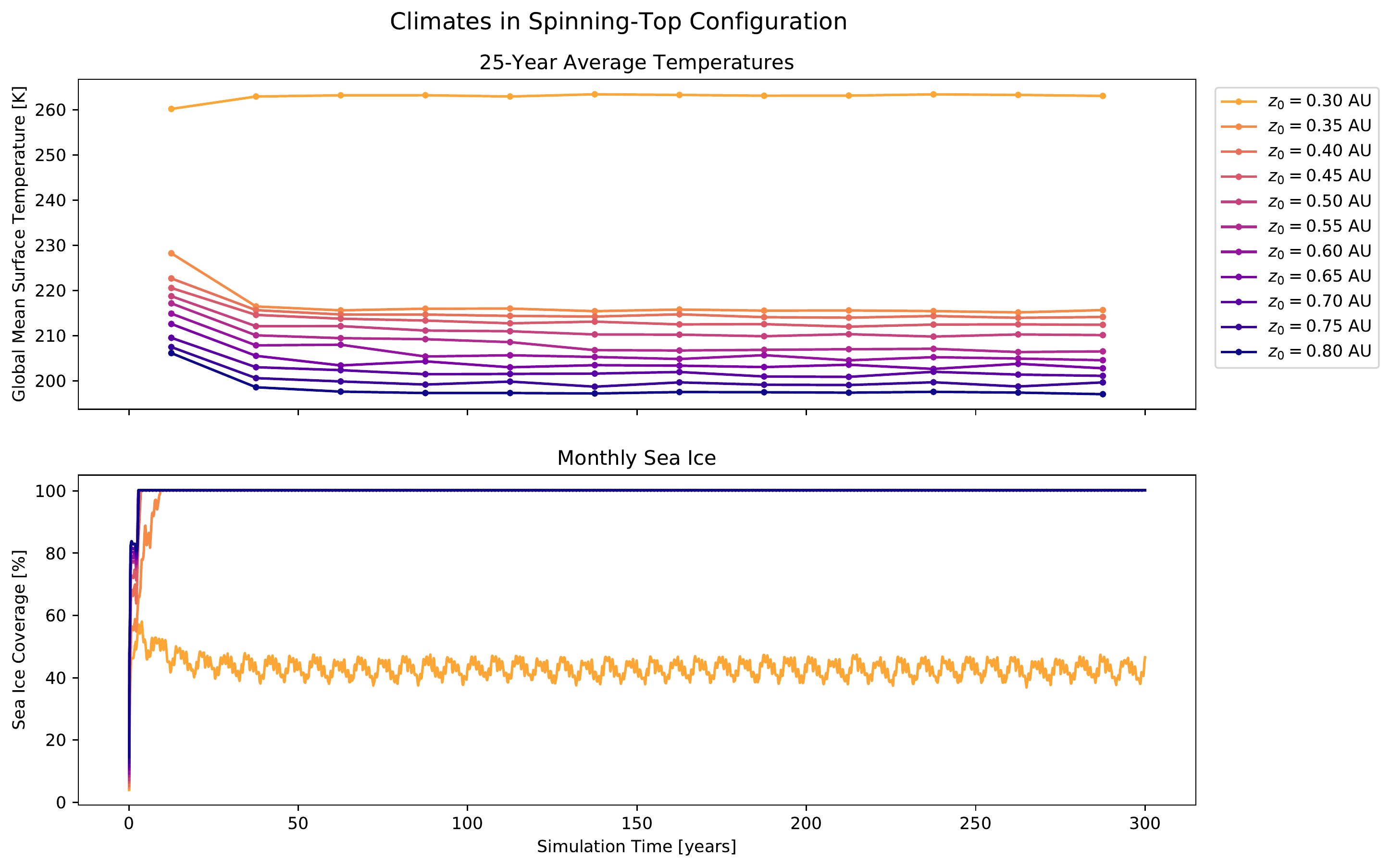}
\caption{Global mean surface temperature averaged over 25-year intervals, and monthly sea ice extent for planets in spinning-top configuration (rotation axis aligned with the direction of motion), for initial elevations above the parent stars' orbital plane ranging from 0.3 AU to 0.8 AU, in 0.5 AU increments. Each planet that goes more than 0.3 AU or so away from the ecliptic plane enters a snowball state, and for our purposes is considered not habitable. Note that 0.3 AU is not a sufficiently large elevation to result in significantly chaotic climates, as seen in the monthly sea ice.\label{fig:spintop}}
\end{figure*}

Assuming a primary insolation of 0.95 L$_\odot$, orbits with initial elevations greater than approximately 0.3 AU result in orbital variations in insolation of more than 600 W/m$^2$. At 0.3 AU elevation and a primary eccentricity of 0.1, planets in the spinning-top configuration experience total darkness for latitudes above 15-18$^\circ$ at maximum excursion. Lower latitudes experience sunlight reduced by $\cos(\psi-\delta)$, where $\psi$ is the latitude and $\delta$ is the declination of the primary relative to the planet's equatorial plane. For example, locations at 45$^\circ$ latitude on the far hemisphere experience a further reduction in sunlight intensity of 45-50\%, for a minimum insolation of 450 W/m$^2$. This is comparable to the intensity of sunlight at similar latitudes during Earth winter. An initial elevation of 0.6 AU, however, results in a minimum insolation of 233 W/m$^2$ at the same latitude. We tested initial elevations ranging from 0.3 to 0.8 AU in 0.5 AU increments, with a 0.95 L$_\odot$ primary and equal-mass black hole, and as shown in \autoref{fig:spintop}, found that all but 0.3 AU resulted in snowball planets, confirming our hypothesis that only planets with initial elevations around 0.3 AU or smaller can be habitable in spinning-top configurations, ruling out this rotation geometry for the chaotic orbits possible at larger initial elevations.

\begin{figure*}
\plotone{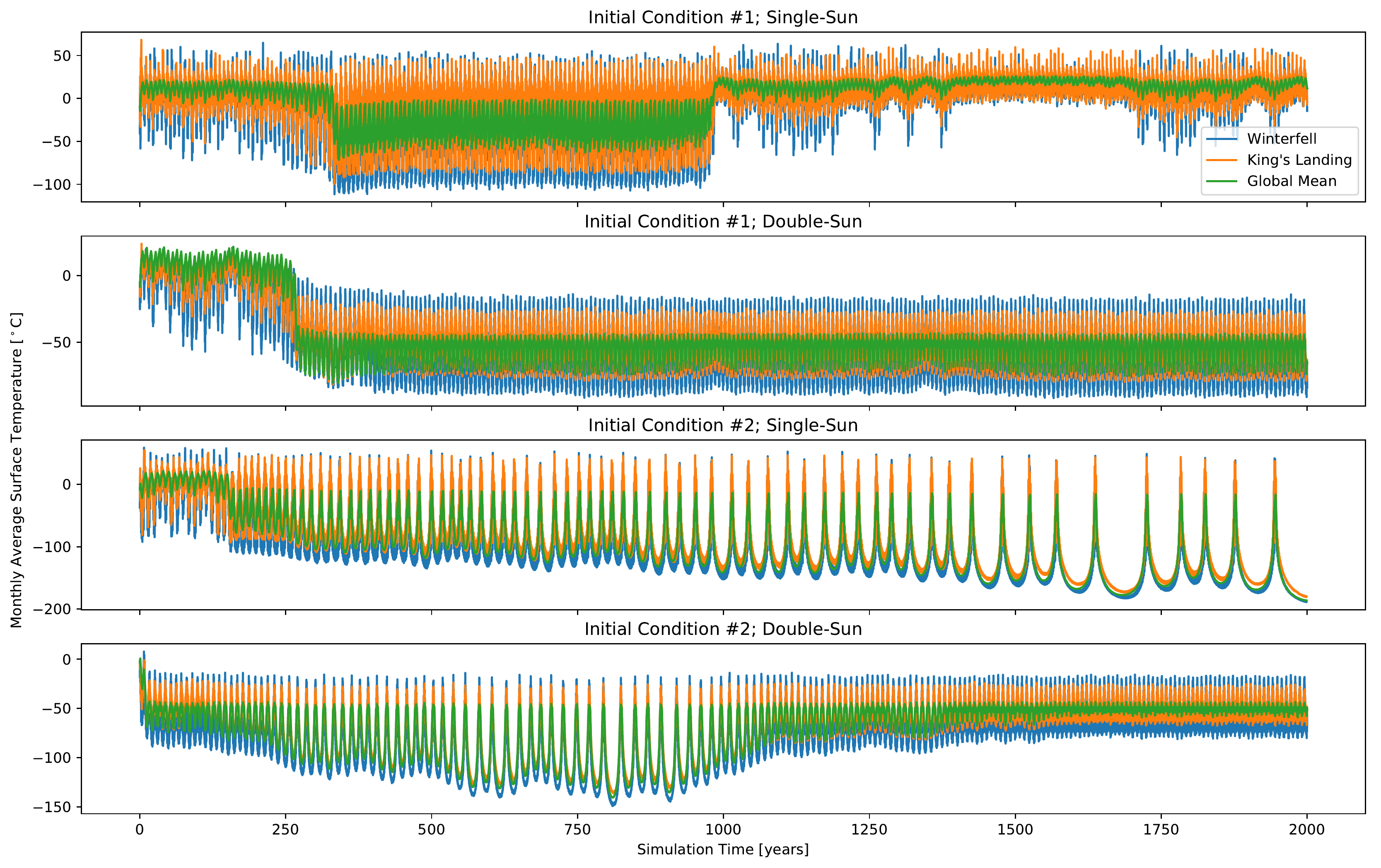}
\caption{Monthly average temperatures averaged across the planet, and at Winterfell and King's Landing, for each of the four 2,000-year simulations. Note that machine precision can result in significant orbital deviations between two simulations with the same initial conditions. In two of our simulations, we do observe the planet both entering and exiting a Long Night, although the change that triggers this in one appears to be smaller than the truly brutal excursions seen in some of the other models, in which decades can pass between crossings.\label{fig:longruns}}
\end{figure*}

With the planet in a tennis-ball spin configuration and initial elevations large enough for chaotic orbits, we find no luminosities for which the planet remains habitable, either due to plunging into model-crashing cold severe enough to chill a White Walker, or due to `pulses' of very high incident sunlight as the planet passes through the primaries' orbital plane. When the primaries are near periapse during this conjunction, incident light is maximized, and in each case where the stellar luminosity is high enough to avoid lethal cold during excursions away from the plane, surface temperatures reach an excess of 100 $^\circ$C during ecliptic crossing.

\begin{figure}
\plotone{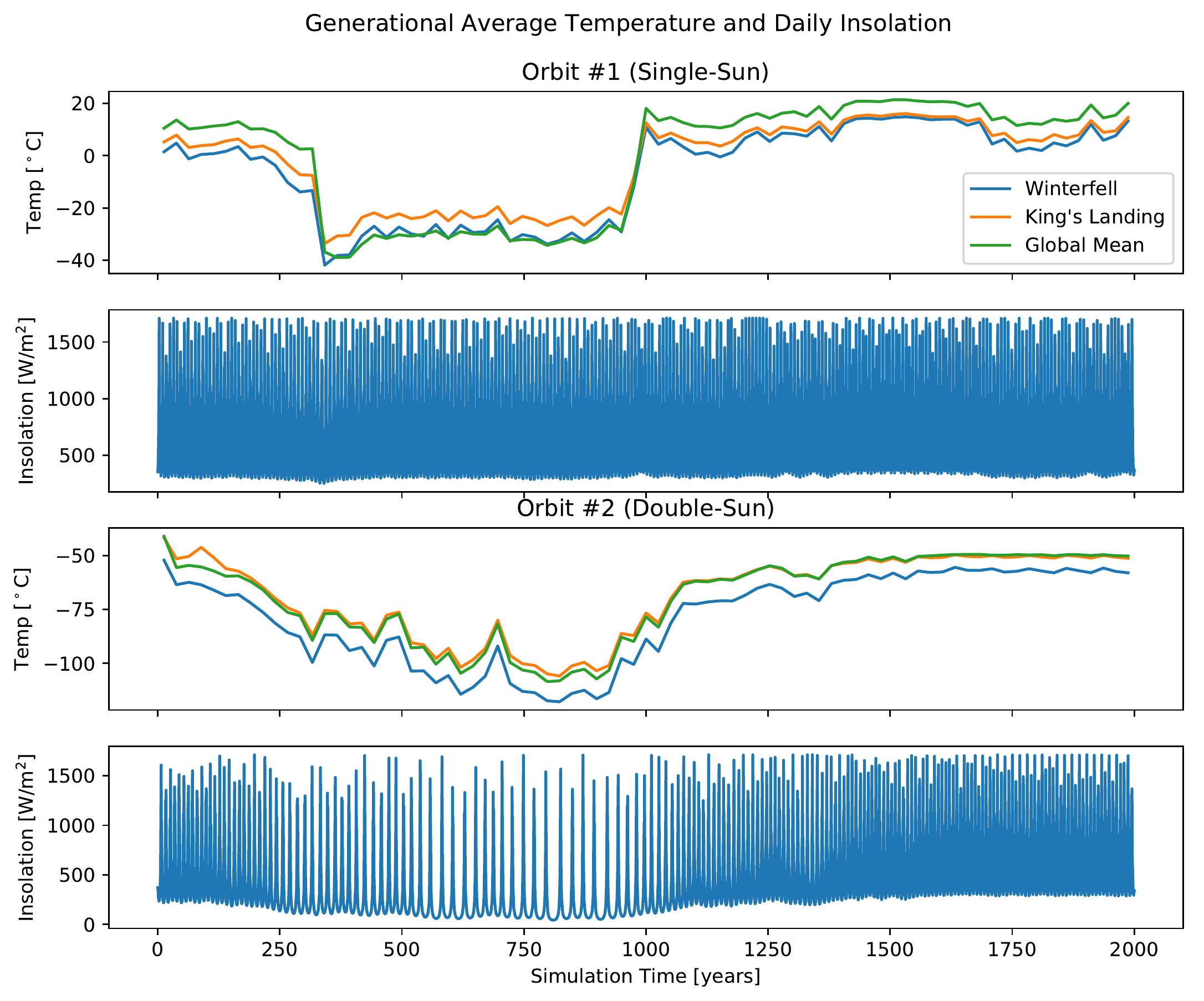}
\caption{Temperature averaged on generation (25-year) baselines, compared to the daily insolation received by the planet for each of the two orbits which produce a Long Night. We note that the Long Night in Orbit \#1 appears to be a snowball episode triggered by a relatively small decrease in insolation, while Orbit \#2 is caused by large excursions to lower insolations.\label{fig:tempinsol}}
\end{figure}

When we artificially place an upper bound on incident flux, we find that we are able to maintain mostly-habitable conditions with tennis-ball configurations at initial elevations high enough to cause significant chaotic variations in excursion distance and duration. We ran four simulations in this configuration, for two different sets of initial conditions: in each case we model both the single-sun case, where one primary is a black hole, and the double-sun case, where both primaries are luminous. In the double-sun case, we set each star's luminosity to be half that of the star in the single-sun case, so that total incident light remains the same. The single-sun case results in overall warmer temperatures, likely due to stronger immediate sunlight reducing daytime sea ice, thereby lowering the planet's overall albedo, increasing the total energy absorbed, and therefore increasing the heat stored for the night. The results are shown in \autoref{fig:longruns}, with the comparison to insolation shown in \autoref{fig:tempinsol}.

\begin{figure}
\plotone{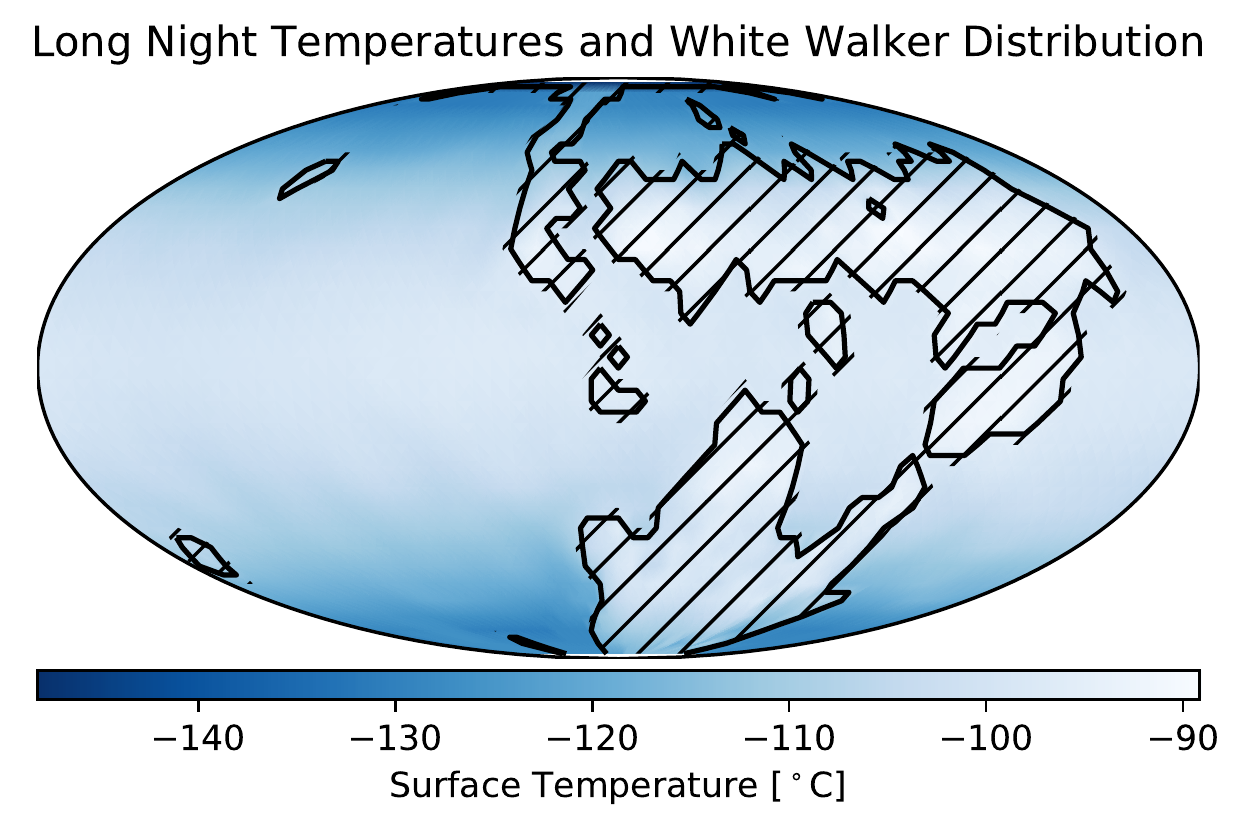}
\caption{Surface Temperatures in the depth of a snowball-like Long Night. Hatched areas indicate lands held by the White Walkers. Note that this would be a catastrophic outcome for the humans of Westeros, and represents one of the more extreme Winters in our simulations.\label{fig:snowball}}
\end{figure}

\begin{figure}
\plotone{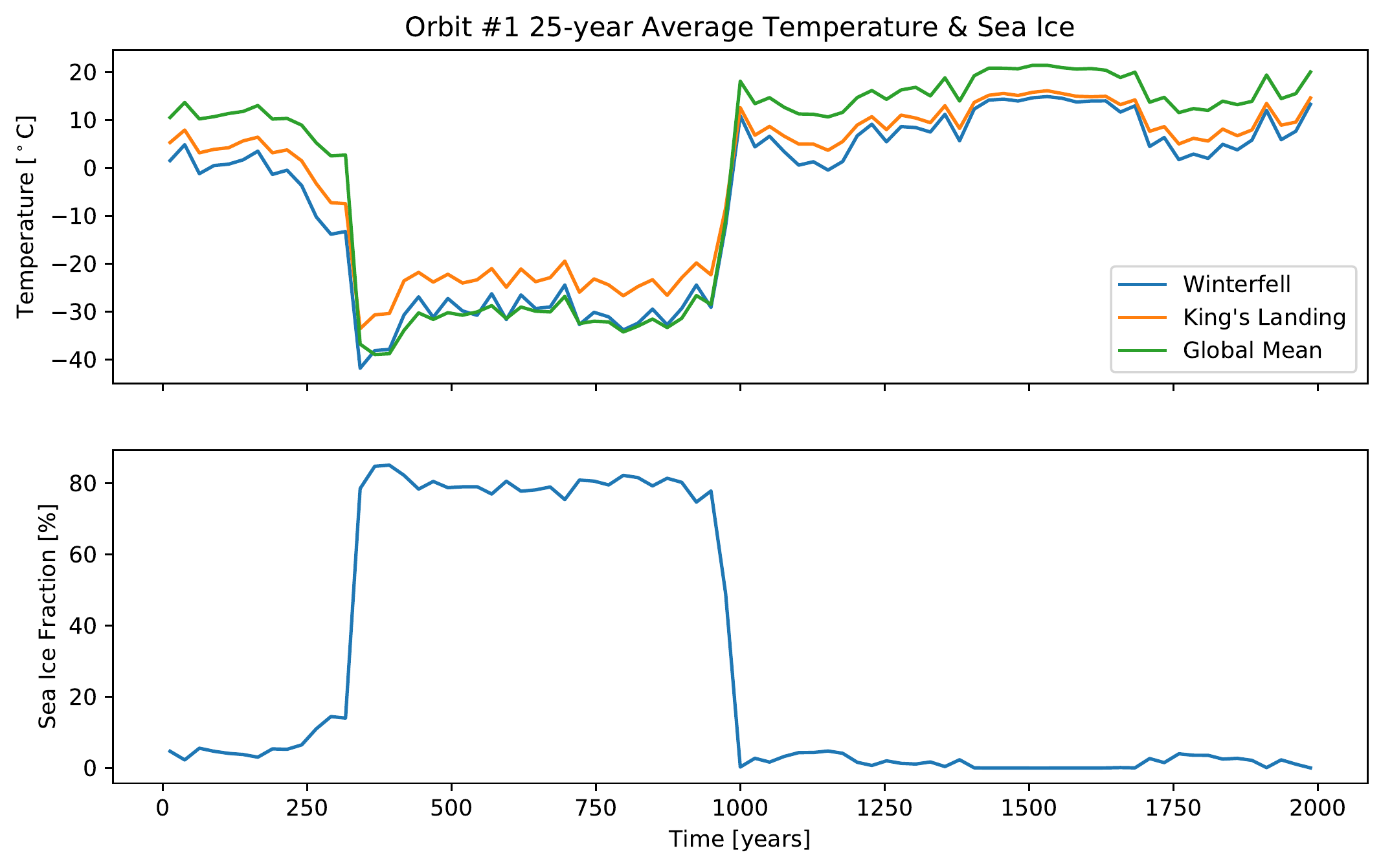}
\caption{25-year average temperatures and sea ice cover for Orbit \#1. The sharp transitions to and from total sea ice cover indicate that the Long Night in this case is a snowball event.\label{fig:snowballwesteros}}
\end{figure}

During Long Nights, the single-sun case reaches above-freezing temperatures during ecliptic crossings, with small reductions in sea ice, but the double-sun case remains in a snowball state for the duration of the Long Night. An example of such a snowball climate is shown in \autoref{fig:snowball}. Looking at 25-year averages, consistent with the experiences of different generations, it does appear that both have characteristics common to snowballs, as shown in \autoref{fig:snowballwesteros}. We find that in both sets of initial conditions, the planet can enjoy centuries of temperate conditions, punctuated at chaotic and random intervals by Winters lasting anywhere from decades to centuries or even millennia between thaws. This suggests that with a means of limiting excess sunlight during ecliptic-crossing, Sitnikov orbits with the planet rotating in a tennis-ball configuration can indeed explain White Walkers.

\section{Discussion}

Our results suggest that Sitnikov orbits can explain the climate of Westeros, but only if specific conditions are met, each of which we will discuss in more detail:
\begin{outline}[enumerate]
    \1 The planet is in a tennis-ball configuration, reaches a sufficiently high elevation above the plane, and the planet has some means of limiting the incident sunlight during the otherwise-deadly crossing of the ecliptic plane.
    \1 The planet is in a spinning-top configuration with small elevation above the plane, but has extremely specific initial conditions such that the number of primary orbits between crossings can be described by a series containing a very large number of consecutive integers of $O(1)$, punctuated by a relatively small number of larger integers \citep{Moser1973}. We note that in the regime of small primary eccentricity and small initial elevation above the plane, the energetics of the system make such an orbit unlikely. 
    \1 Some other element of the planet, such as its axial tilt, also varies chaotically.
\end{outline}

\subsection{Limiting sunlight during ecliptic-crossing}
We have found that in order for planets to survive in more-chaotic orbits with greater elevation above the plane without lethally cold temperatures every few months, they must experience lethally hot temperatures during crossing of the ecliptic plane. This can be prevented if the planet has some means of limiting the sunlight reaching the planet. One reasonably-feasible solution might be a space-based megastructure, such as an orbital sunshade, which has been proposed for climate engineering on Earth \citep{Early1989,Angel2006}. This megastructure could take the form of a large swarm of smaller sunshades which orbit the barycenter at various inclinations and semimajor axes to shield the planet throughout the period of greatest danger, or they could orbit the planet and re-orient themselves to let sunlight through during excursions away from the plane. Conversely, if stellar luminosity is low enough to survive ecliptic-crossing, large space-based mirrors could redirect additional sunlight to the planet during excursions, keeping the planet habitable.

Proposing that Westeros is accompanied by orbital megastructures may sound like an outlandish departure from known Westerosi history, but we argue that this is actually entirely consistent with what is known about the history of Westeros's inhabitants. The Wall, standing 700-900 feet tall and stretching for 300 miles, is described by the Westerosi as imbued with powerful magical spells that suffice to keep White Walkers at bay \citep{georgemartin2003}. While the Wall itself was built by Bran the Builder, the supposed `magical' properties were provided by the Children of the Forest \citep{georgemartin2014}, a race of beings described as possessing powerful magic. We note here that ``any sufficiently advanced technology is indistinguishable from magic" \citep{arthurclarke1973}, which suggests that it is quite possible that the Children of the Forest were in fact an extremely technologically-advanced race, capable of large-scale engineering and therefore certainly able to construct orbital megastructures. We note however that that this shifts the balance in favor of higher luminosity and sun shades rather than lower luminosity and sun mirrors, as the latter would suggest that the White Walkers could always be easily and permanently defeated by increasing the intensity of redirected sunlight, and would preclude the necessity of experiencing a Long Night. Alternatively, we note that the Long Night in our simulations lasts much longer than a single generation, so perhaps the Children of the Forest did intervene in this way to drive back the White Walkers. If this is the case, then Jon Snow and the humans of Westeros are surely doomed in the coming Winter.

\subsection{A chaotic spinning-top configuration}

As we have noted, orbits with small initial elevations and small eccentricity tend to be more stable than orbits with higher primary eccentricity or higher initial elevation, which makes them poor candidates for triggering the Long Night as frequently as every 8,000 years. Nonetheless, a spinning-top configuration with small eccentricity and lower initial elevation is attractive for characterizing Westero's orbital dynamics, as Westerosi seasons are mild enough to be Earth-like, and the Long Night is described as not just cold, but dark---as if the planet suddenly traveled far enough above the plane for the primaries to sink below the horizon at Westerosi latitudes for a generation. In our tennis-ball configuration, we must accept that sometimes the declination of the Sun(s) will change dramatically, particularly during ecliptic crossing, and that the `darkness' of the Long Night is merely metaphorical, or in reference to the reduced intensity of sunlight. As \citet{Freistetter2018} point out, Moser's Theorem does posit that a solution should exist which mixes the more-appealing diurnal characteristics of the spinning-top configuration with the dramatic Long Night excursions of our more-chaotic orbits \citep{Moser1973}. However, large excursions are the consequence of the planet receiving a large amount of momentum from near-periapse primaries during a crossing, and then experiencing comparatively less deceleration at maximum excursion due to the primaries moving to apoapse. In the case of low eccentricity and low elevation, the size of any individual `kick' is necessarily small, and a very large number of orbits are required to build up significant momentum, and that build-up ought to be gradual. Moser's Theorem states that this must not necessarily be the case, and sudden excursions should be possible, but we argue that these energetics suggest it is at the very least unlikely, occupying a small portion of the phase space. 

\subsection{Chaotic variations in axial tilt}

A third possibility which requires neither orbital megastructures nor unlikely orbital configurations is that the planet's axial tilt is in neither the spinning-top nor tennis-ball configuration, but instead varies chaotically. In a system where the planet's angular momentum vector is mis-aligned with the primaries' angular momentum vector, the orientation of the planet's rotation axis will precess chaotically. The presence of a moon can stabilize this precession to some extent, but if the planet's orbital characteristics vary in time, then that variation and torques from other planets or nearby stars can serve to destabilize the rotation axis \citep{Vinson2017,Makarov2018,Saillenfest2019}. A major Westerosi religion invokes `the light of The Seven', which could be understood to refer to 7 other planets in the system, which may be large, close, or both \citep{Martinbook}. It is therefore likely that a planet could be in a low-eccentricity, low-elevation configuration with an axial tilt similar to the spinning-top configuration, and then occasionally chaotically precess to a very different axial tilt, moving Westeros from temperate latitudes to polar, darker latitudes, especially if the new tilt points Westeros away from a single luminous primary's apoapse, similar to how the Arctic Circle on Earth is plunged into months of darkness during Northern Hemispheric winter. Further work will be needed to model the effect of such an additional source of chaos. 

\subsection{Other alternatives}

While we find that with some caveats, Sitnikov orbits could be the correct explanation, there are still other explanations that could work as well. We do not believe that \citet{Kostov2013}'s circumbinary proposal can yet be dismissed. The non-double-sun problem, as we have seen, is easily solved by invoking a black hole as one of the primaries. In addition, other studies have shown that chaos in circumbinary systems can be significant \citep{Batygin2015}, and the climatic consequences non-negligible \citep{May2016,Popp2017}. We recommend further work to identify parts of the chaotic circumbinary parameter space which could replicate the Long Night phenomenon. Furthermore, we still cannot rule out the role that magic may play in bringing about White Walkers.

\subsection{Model Caveats}

Our model comes with a few caveats. Because our machine-level precision is finite, and Sitnikov orbits are chaotic, the orbital solutions represented by our simulations may differ from reality, as we are only approximating positions and velocities at each timestep. Additionally, PlaSim does not include oceanic heat transport through ocean currents \citep{Fraedrich2005}, which could significantly change the habitability and climate dynamics of the planet \citep{Yang2019}. We also note that our land distribution is approximate and speculative.

\section{Conclusion}

We have used a modified 3D climate model to simulate the climate of Westeros assuming a Sitnikov orbital configuration, following the proposal in \cite{Freistetter2018}. We find that there do exist configurations which result in both habitability and long, cold periods characteristic of the Long Night. We note however that this explanation is substantially less likely if the Children of the Forest have not constructed orbital megastructures. Introducing chaotic variations in axial tilt may make a Sitnikov explanation more likely, but we argue that a chaotic, vanilla P-type circumbinary orbit cannot be ruled out.

Furthermore, we note that some of our models resulted in transitions to snowball climates. This suggests that chaotic variations in orbital characteristics or axial tilt, particularly in unusual planetary systems, could serve as a trigger for snowball glaciation on terrestrial exoplanets whose orbital environment differs significantly from Earth's. For these planets, we predict that Winter is Coming.

\section*{Acknowledgments}

We would like to thank George R.R. Martin for crafting such an inspiring yet confusing world, and HBO for inspiring us with amazing visuals. We would also like to extend gratitude to the system administrators of our local computing cluster, for not noticing (or turning a blind eye to) this egregious misuse of resources. J.J. Zanazzi provided valuable comments on the evolution of a planet's spin axis orientation in mis-aligned systems. Michael B. Lund, editor of \textit{Acta Prima Aprilia}, provided advice on submission guidelines and deadlines. Ariel Amaral provided necessary emotional support and enthusiasm. We acknowledge that this work could not have been performed without a large amount of resources produced by the Muskoka Roastery Coffee Co. This work was performed during the authors' free time and, beyond computing resources, did not make use of public tax dollars in the form of any kind of grant money.

We would furthermore like to acknowledge that our work was performed on land traditionally inhabited by the Wendat, the Anishnaabeg, Haudenosaunee, M\'{e}tis, and the Mississaugas of the Credit First Nation.

\appendix

\section{Stellar Angles}\label{appendix:angles}
\begin{itemize}
\item In the tennis-ball configuration, the hour angle of one of the primaries in Westeros's sky is 0 or $\pi$ when the planet's elevation above the plane $z(t)=0$, and when $z(t)\neq0$, is given by 
\begin{equation}
    h(t)=\tan^{-1}\left(\frac{r(t)\cos\theta(t)}{z(t)}\right)-\frac{\pi}{2}
\end{equation} 
\item $\theta(t) = \nu(t)-\pi$ for the western primary, and $\theta(t) = \nu(t)$ for the eastern primary.

\item The primary's declination is given by 
\begin{equation}
    \delta(t)=\tan^{-1}\left(\frac{r(t)\sin\theta(t)}{\sqrt{z^2(t)+r^2(t)\cos^2\theta(t)}}\right).
\end{equation}
\item In the spinning-top configuration, the hour angle is given by \begin{equation}
    h(t)=\pi-\theta(t)
\end{equation} and the declination is given by
\begin{equation}
    \delta(t) = \frac{z(t)}{|z(t)|}\left[\frac{\pi}{2}-\left|\,\tan^{-1}\left(\frac{r(t)}{z(t)}\right)\right|\right]
\end{equation}
\item A local hour angle $h'(t)$ is computed from $h(t)$ by adding the planet's angular rotation rate, $\Omega$, and local longitude $\varphi$ such that
\begin{equation}
    h'(t) = h(t) + \Omega{t} + \varphi
\end{equation}
\item With these angles, the cosine of the stellar zenith angle $\Phi_z$ can be calculated for any latitude $\psi$ and longitude $\varphi$ with 
\begin{equation}
    \cos\Phi_z = \sin\psi\sin\delta(t) + \cos\psi\cos\delta\cos{h'(t)}
\end{equation}
\item Hodor Hodor, Hodor Hodor Hodor, Hodor:
\begin{equation}
    \text{Hodor} = \text{Hodor}
\end{equation}
\end{itemize}

\end{document}